\documentclass[aps,prl,twocolumn,superscriptaddress]{revtex4-1}

\usepackage{graphicx}
\usepackage{dcolumn}
\usepackage{bm}
\usepackage{mathtools}
\usepackage{siunitx}
\usepackage{csquotes}

\usepackage[unicode=true,
bookmarks=true,bookmarksnumbered=true,bookmarksopen=true,bookmarksopenlevel=2,
breaklinks=false,pdfborder={0 0 1},backref=false,colorlinks=true]
{hyperref}
\hypersetup{pdftitle={Optical trapping and enantiomeric recognition of single chiral nanoparticles},
	pdfauthor={Schnoering Gabriel et al.},
	pdfsubject={Optical trapping},
	pdfkeywords={Schnoering, Gabriel, ISIS, chiral, optical trapping,},
	linkcolor=black, citecolor=blue, urlcolor=blue, filecolor=blue, pdfpagelayout=OneColumn,
	pdfnewwindow=true, pdfstartview=XYZ, plainpages=false}

\begin{document}

\title{Three-dimensional enantiomeric recognition of optically trapped single chiral nanoparticles}

\author{Gabriel Schnoering}
\affiliation{ISIS \& icFRC, University of Strasbourg and CNRS, 8 all\'{e}e Gaspard Monge, 67000 Strasbourg, France.}
\author{Lisa V. Poulikakos}
\affiliation{Optical Materials Engineering Laboratory, ETH Z{\"u}rich, 8092 Z{\"u}rich, Switzerland}
\author{Yoseline Rosales-Cabara}
\affiliation{ISIS \& icFRC, University of Strasbourg and CNRS, 8 all\'{e}e Gaspard Monge, 67000 Strasbourg, France.}
\author{Antoine Canaguier-Durand}
\affiliation{Laboratoire Kastler Brossel, Sorbonne Universit\'e, CNRS, ENS-PSL University, Coll\`ege de France, Paris, France.}
\author{David J. Norris}
\affiliation{Optical Materials Engineering Laboratory, ETH Z{\"u}rich, 8092 Z{\"u}rich, Switzerland}
\author{Cyriaque Genet}
\email[Electronic address: ]{genet@unistra.fr}
\affiliation{ISIS \& icFRC, University of Strasbourg and CNRS, 8 all\'{e}e Gaspard Monge, 67000 Strasbourg, France.}

\date{\today}

\begin{abstract}
We optically trap freestanding single metallic chiral nanoparticles using a standing-wave optical tweezer. We also incorporate within the trap a polarimetric setup that allows to perform \textit{in situ} chiral recognition of single enantiomers. This is done by measuring the $S_3$ component of the Stokes vector of a light beam scattered off the trapped nanoparticle in the forward direction. This unique combination of optical trapping and chiral recognition, all implemented within a single setup, opens new perspectives towards the control, recognition, and manipulation of chiral objects at nanometer scales.
\end{abstract}

\maketitle

Artifical chiral nanostructures have opened new perspectives in the field of colloidal science, optics, and spectroscopy \cite{BenMoshe2013,valev2013chirality,boriskina2014singular}. Chiral plasmonic nanostructures, for instance, have led to the possibility of enhancing chiroptical signals through the excitation of so-called superchiral electromagnetic fields \cite{tang2010optical,tullius2015superchiral,alizadeh2015enhanced,mcpeak2015ultraviolet,schaferling2016chiral}. New strategies proposed recently for preparing colloidal suspensions of chiral nano-objects have allowed fascinating experiments on active Brownian motions and have sparked strong activities \cite{meinhardt2012separation,nourhani2015guiding,schamel2013chiral,Mcpeak2014,yeom2015chiral,Bechinger2016RMP}. 
In parallel, chiral structures interact specifically with chiral light fields, as seen in particular through the emergence of new types of optical forces recently described \cite{li2010theory,guzatov2011chiral,canaguier2013mechanical,cameron2014discriminatory,wang2014lateral,dionne2017nano}. While such forces have only been probed experimentally at the micrometer scale \cite{cipparrone2011chiral,tkachenko2014optofluidic}, the outstanding chiroptical signatures associated with these new artificial chiral nano-objects could facilitate, despite their small sizes, the observation of such chiral optical forces on nanometer-sized objects. 
 
To move toward manipulating chiral matter at nanometer scales, one crucial step is the spatial control of a single chiral nano-object in freestanding conditions. In this Letter, we develop an optical tweezer capable of trapping single chiral metallic nano-objects that diffuse in a fluidic cell. We demonstrate three-dimensional stable optical trapping of single artificial Au nanopyramids (NPys) in both enantiomeric forms. Simultaneously, the enantiomeric form of the trapped NPy is recognized through a far-field polarization analysis of the scattered light inside the trap, at the single-particle level. Our experimental strategy is grounded on fundamental concepts (conservation law of optical chirality and chiral scattering) that lead to new physical discussions, as exemplified in our use of chiral symmetries in the context of optical trapping. Our work shows how such concepts, which are at the core of many current debates and discussions, can turn operational in the experimental study of chiral matter at the nanoscale. In particular, the conservation law of optical chirality \cite{poulikakos2016optical,Nienhuis2016conservation} enables the novel physical concept for single-particle enantiomeric recognition presented in this work.

\begin{figure}[ht!]
  \includegraphics[width=0.4\textwidth]{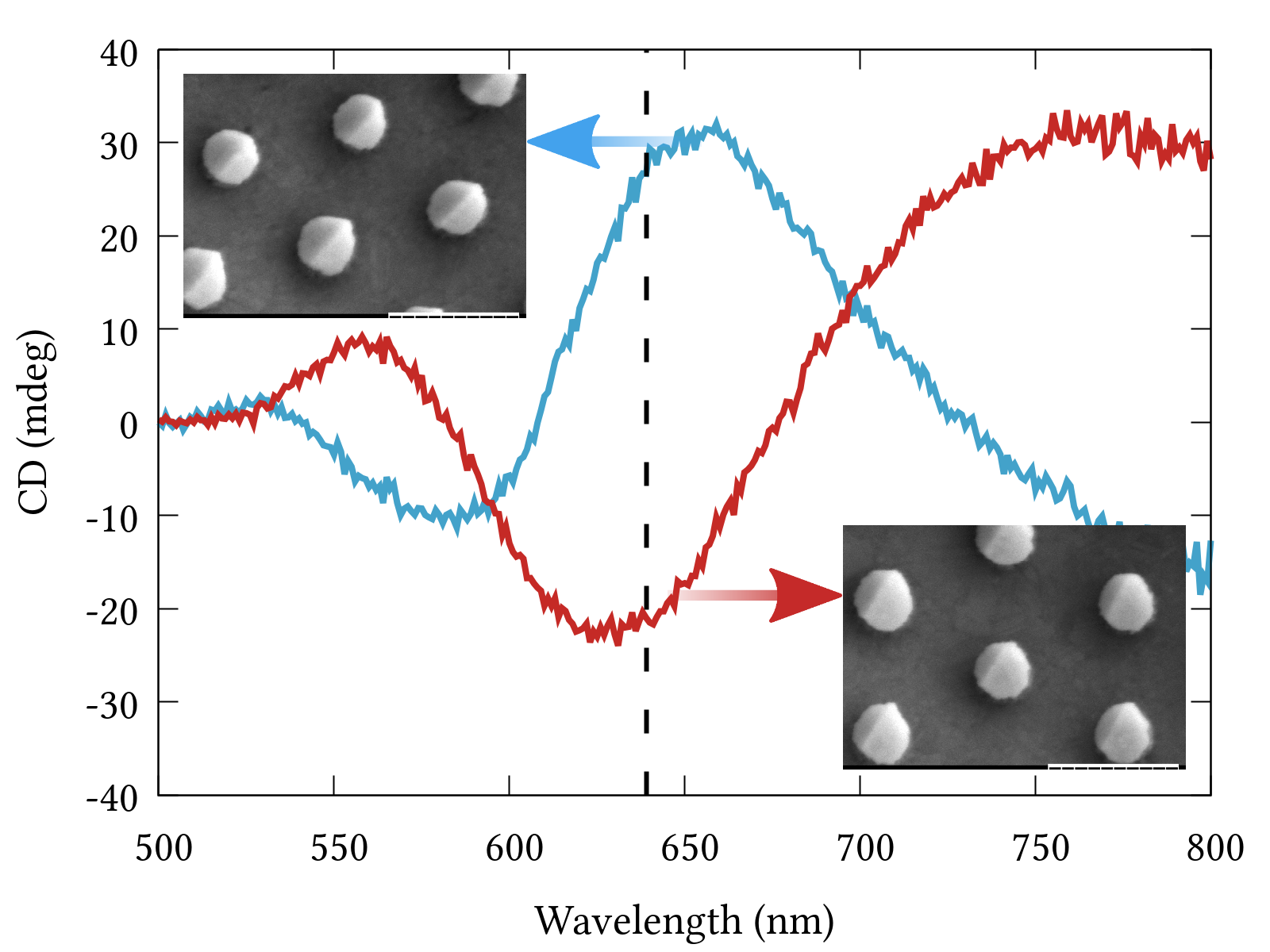}
  \caption{Circular dichroism (CD) spectra measured through $1$ cm thick cuvette for right-handed (red) and left-handed (blue) NPy dispersions, each prepared in 300 $\mu$L of a buffer made of $15$ mL $0.1$ M trisodium citrate dihydrate and $100~\mu$L $0.1$ M citric acid (pH$=7.32$). The dashed line represents the wavelength of the probe laser used in our experiment. The corresponding SEM images of the NPys, taken directly after lift-off, are displayed as insets for the left-handed (left upper corner) and right-handed (right lower corner) NPys. The scale bars are 500 nm.}
 \label{fig:CDchiral}
\end{figure}

Colloidal dispersions of Au chiral nanopyramids (NPys) fabricated via high-index off-cut Si wafers -see \cite{Mcpeak2014} for all details- are prepared in stabilized solutions with typical NPy sizes of the order of $150$ nm. The insets in
Fig. \ref{fig:CDchiral} show SEM images of such chiral NPys after being stripped off from the Si template, for left and right handeness, respectively. A surprisingly strong circular dichroism (CD) signal is measured on these objects, as seen in Fig. \ref{fig:CDchiral} \cite{Mcpeak2014}. The CD spectra show a clear sign inversion between the two opposite enantiomeric forms of the NPys. Importantly for the experiments, the CD response of the chiral NPys peaks at $639$ nm, a wavelength at which a second laser can easily be tuned to and exploited for the chiral recognition protocol described below.

\begin{figure}[ht!]
  \centering
  \includegraphics[width=0.35\textwidth]{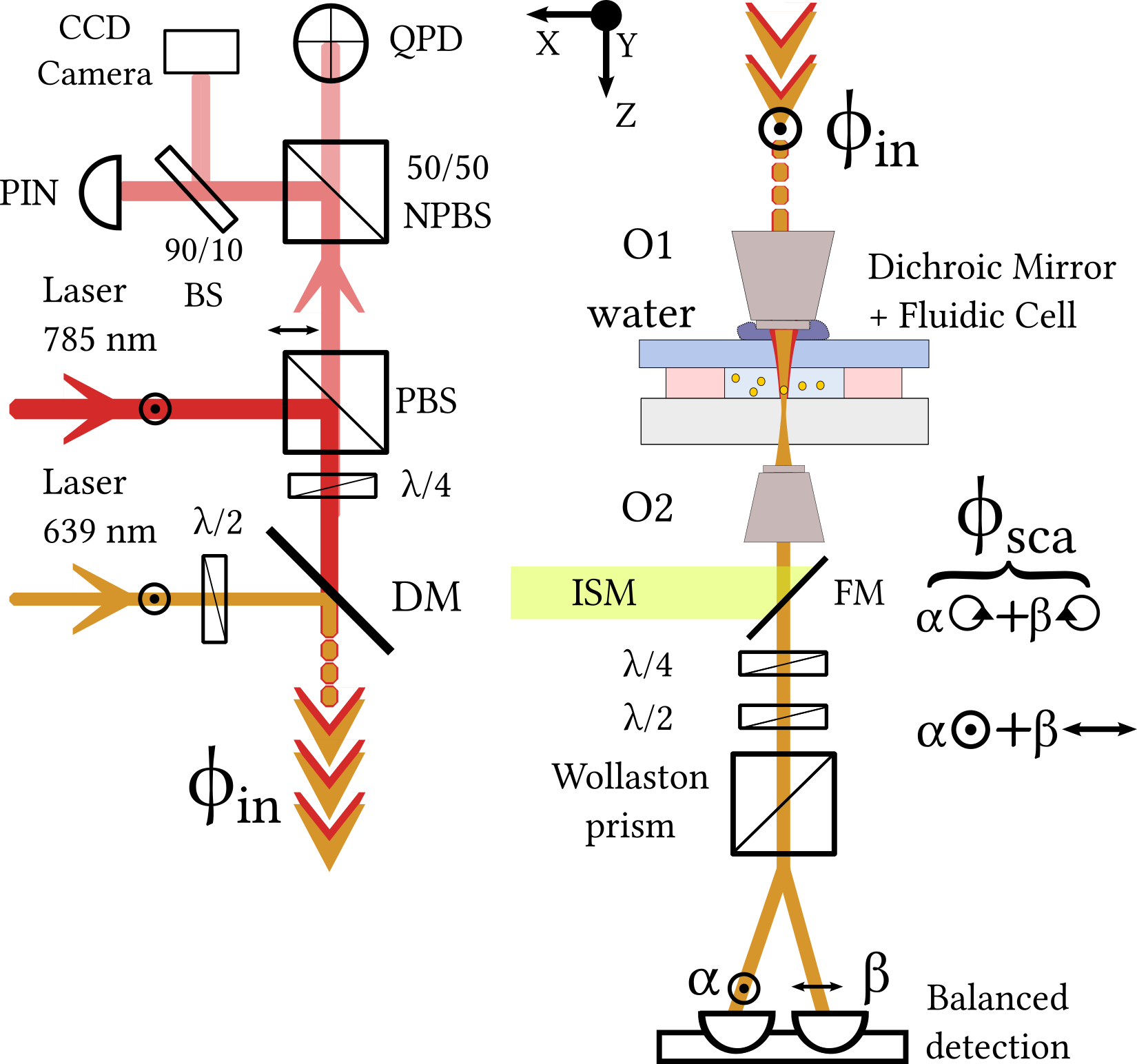}
  \caption{Scheme of the experimental setup used for trapping a single NPy and for performing chiral recognition on it. The standing-wave optical trap consists of a circularly polarized TEM$_{00}$ beam from a 785 nm diode-laser (45 mW) sent into a water immersion objective (O1, $60\times$, $1.2$ numerical aperture (NA)) and focused in a water cell (deionized water, $120~\mu$m thick). The beam is reflected by a dichroic mirror (the end-wall of the fluidic cell) placed at a distance ca. $3~\mu$m from the beam waist, creating a standing wave pattern within which a single NPy can be trapped. The chiral recognition setup involves a linearly polarized $\phi_{\rm in}$ probe laser at $639$ nm (100 $\mu$W) injected inside the trap with a $45^\circ$ dichroic mirror (DM) and sent through the trap using a dichroic end-mirror. The polarization analysis is performed behind a second (collection) objective (O2, NA 0.6, $40\times$) on the interfering signal between $\phi_{\rm in}$ and the field $\phi_{\rm sca}$ scattered by the single trapped NPy with a $\lambda / 4$ quarter-wave plate at $45^\circ$, followed by a $\lambda / 2$ half-wave plate, and a Wollaston prism. The Wollaston prism separates the incident beam into two linear (vertical and horizontal) polarized beams by an angle of $20^\circ$. Both output channels are then sent to a balanced photodetector. The low-power laser beam ($594$ nm, green trace) for the interferometric scattering microscopy is injected with a flip mirror FM, and therefore available throughout the experimental session.}
  \label{fig.trap}
\end{figure}

With diameters of ca. $150$ nm, the NPys are metallic particles that cannot easily be trapped in three dimensions using a conventional optical tweezer approach \cite{svoboda1994,hansen2005}. To reach good trapping conditions, we built a standing-wave optical trap (SWOT). This enables the axial immobilization of a single metallic nanoparticle at an anti-node of the standing-wave pattern created by the reflection of the trapping laser ($\lambda_T=785$ nm) beam on a mirror placed at a given distance from the beam waist, as depicted in Fig. \ref{fig.trap} \cite{schnoering2015}. For the transverse confinement, the compensation between the Poynting vectors of the incoming and reflected beams leads to a strong reduction of the axial scattering force that can be easily overcome by the gradient force induced by the focusing effect of the objective. The combination of axial and transverse confinements leads to the three-dimensional trapping of the metallic nanoparticle. In such a counter-propagating beam configuration, the scattering forces induced on the nanoparticle therefore stabilize the trap \cite{zemanek1999optical}. This is in clear contrast with conventional single-beam optical traps where the scattering forces tend to push away from the waist any metallic particle of size larger than 100 nm. In addition, as discussed in \cite{zemanek2015nonspherical} for instance, the pyramidal anisotropic shape of the NPy is expected to even further enhance the trapping efficiency.

\begin{figure}[ht!]
  \includegraphics[width=0.4\textwidth]{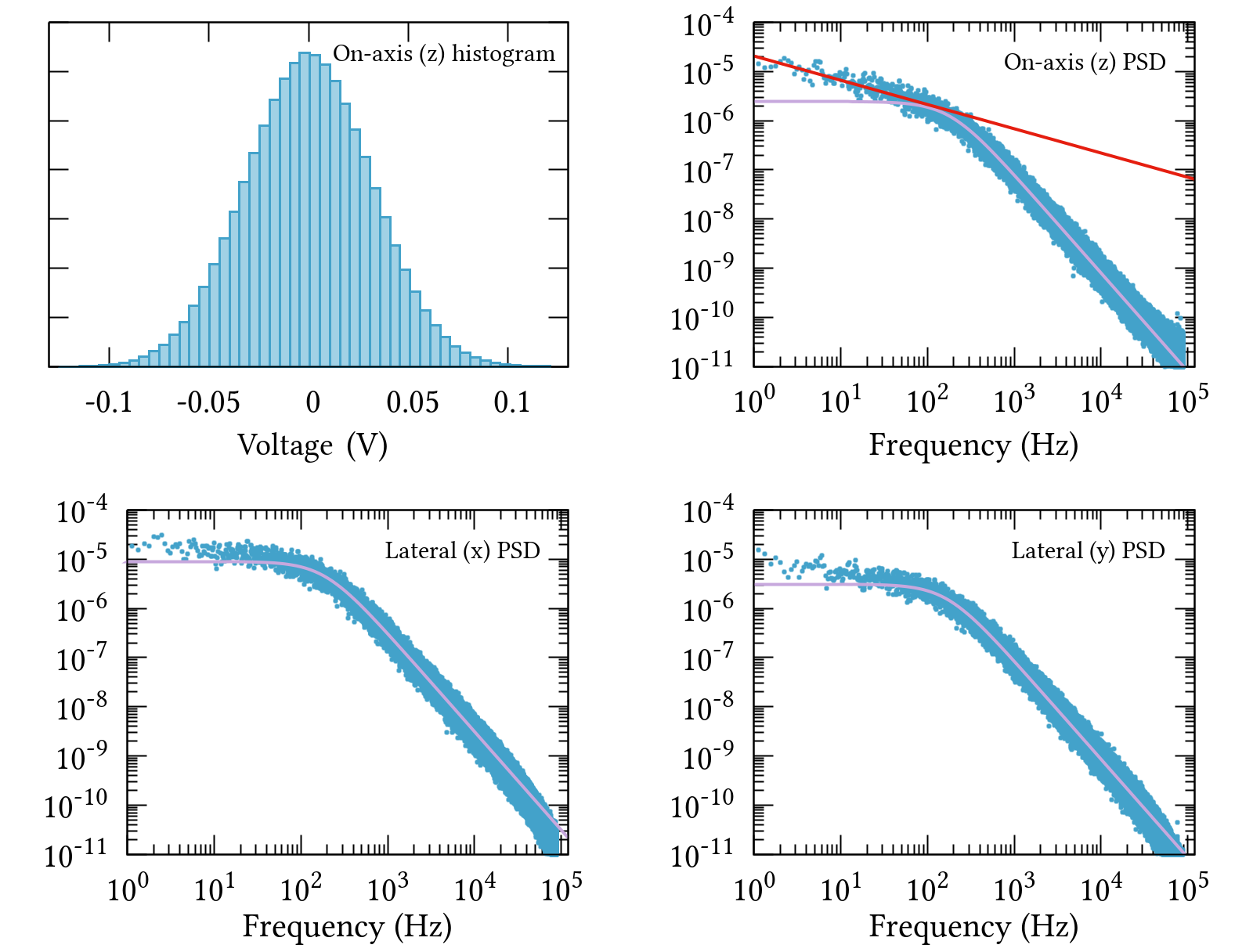}
  \caption{(a) The intensity histogram along the optical axis shows a Gaussian like distribution of positions. Such behavior is expected for the motion of a single nano-object within a stable optical trap. (b-d) Power spectral densities (PSD) acquired for 36 seconds along the 3 $x,y,z$ spatial axis for a trapped chiral NPy. A PSD in blue color is $8\times$ averaged  and the continuous purple curve gives the best Lorentzian fit. (b) PSD along the optical $z$ axis, calculated directly from the intensity of the trapping beam scattered back into the objective and recorded with the PIN photodiode (see Fig. \ref{fig.trap}). The red line gives the best fit of the data on the low frequency part of the spectrum (from 1 to 200 Hz) and has a slope of $-0.49$. (c)-(d) Transverse horizontal $x$ / vertical $y$ signal components are acquired by the quadrant photodiode (QPD -Fig. \ref{fig.trap}). They both show an almost Lorentzian shape, signature of a harmonic trapping potential.}
 \label{fig.chiral.trap}
\end{figure}

For our experiments, colloidal dispersions of NPys are enclosed in a fluidic cell 120 $\mu$m thick. We reduced any electrostatic effects as much as possible by negatively charging the NPys using a citrate buffer solution and the surface of the SWOT end-mirror dip-coated for 5 minutes in a 5 wt \% polystyrene sulfonate solution. 
The main constraint for our experiments is the necessity to work with very dilute dispersions appropriate when trapping single NPys. But despite the careful choice of the buffer, the quality of the dispersion evolves in time. NPys tend to adhere on the walls of the fluidic cell, reducing the concentration of the dispersion to levels that can not be exploited experimentally.  This unavoidable effect puts stringent constraints on the time available for repeated experiments on different objects within the same dispersion. In addition, the large surface-to-volume ratio of each NPys leads to the formation of aggregates, in numbers that increase with time. Such aggregates are more likely to be trapped than single NPys and therefore demand a capacity to discriminate between single and aggregated objects. To reach this level of control, we have implemented an interferometric scattering microscopy (ISM) \cite{Lindfors2004}. The setup is described in Supplemental Material (Sec. A) and the method ensures that our experiments involve single NPys, keeping trapped objects only associated with the smallest ISM signals and with diffusive behavior similar to those recorded for single $150$ nm Au nanospheres.

Trapping single chiral NPy in stable conditions is then done by carefully positioning the end-mirror of our SWOT at a distance of $2~\mu$m behind the waist of the trapping beam. This stability is clearly observed on the power spectral density (PSD) associated with the motion of the trapped NPy in each of the three dimensions displayed in Fig. \ref{fig.chiral.trap}. The Gaussian distribution of the fluctuations in the intensity of the recollected trapping beam, as measured with the PIN photodiode, clearly demonstrate that the single NPy is well localized in space, i.e. well trapped. 

Above the roll-off frequency of the optical trap, the PSDs also precisely match the $f^{-2}$ signature of free Brownian motion. This implies that at such high frequencies, the NPy freely diffuses inside the optical trap. Nevertheless, a closer look at the PSDs reveals an interesting deviation below the trap roll-off frequency. The measured PSD indeed departs from the Lorentzian PSD profile expected in the case of a simple harmonic optical trap. This is particularly clear along the optical axis in Fig. \ref{fig.chiral.trap} (a) with a spectral dependence $S_z[f]\propto f^{\alpha}$ ($\alpha \simeq -0.49$). While this low-frequency power law can slightly depend on the position of the end-mirror when positioned at a distance of about $2~\mu$m behind the waist of the trapping beam, we have checked that we could never record the low-frequency $k_{\rm B} T / \eta \pi^2$ plateau expected for a Lorentzian PSD measured for a viscous drag $\eta$ at temperature $T$. In fact, the $\sim -0.5$ exponent is the largest exponent we measured, which, remarkably, corresponds to the best trapping conditions for our SWOT. It is beyond the scope of this work to understand exactly this low-frequency deviation. But considering that it lies between the Lorentzian plateau and the $1/f$ shot noise spectral signature, we anticipate that the uneven facets of the NPy diffusing within a limited trap volume (ca. $0.01~\mu$m$^3$) could be the source for such additional, broadly distributed, correlations in the low-frequency part of the measured scattering signal. 

We now exploit a fundamental consequence of the conservation law of optical chirality \cite{poulikakos2016optical,Nienhuis2016conservation}. Upon achiral excitation, a lossy, dispersive chiral object selectively dissipates optical chirality and must therefore break, in the scattering, the initial balance in left ($\sigma_L$) vs. right ($\sigma_R$) circular polarizations of the excitation field. This unbalanced scattering is determined in direct relation with the chiral nature of the scattering object, hence its enantiomeric form. This relation leads to design an enantionmeric recognition protocol that we now describe.

We first model the scattering on a single NPy using simple paraxial circularly dichroic Jones matrices $J^{\pm}$ associated with each of the two $\pm$ NPy enantiomers with 
\begin{eqnarray}
J^+ &=&
\left(\begin{array}{cc}
\alpha & 0\\
0 & \beta
\end{array}\right) \label{Jp} \\  
J^- &=&
\left(\begin{array}{cc}
\beta & 0\\
0 & \alpha
\end{array}\right) \label{Jm}
\end{eqnarray}
written in the basis of the circularly polarized states $(\sigma_L,\sigma_R)$. In this formulation of purely circularly dichroic nano-objects, the absence of mirror symmetry that characterizes the NPy chirality simply corresponds to real $\alpha\neq\beta$ parameters \cite{drezet2014reciprocity}. 

Illuminated by an incident field linearly polarized $\phi_{\rm in} = \frac{1}{\sqrt{2}}(\sigma_L +\sigma_R)$, the NPy scatters a field $\phi_{\rm sca} ^{\pm} = J^{\pm}\phi_{\rm in}$ in the forward direction with non-equal ($\alpha, \beta$) weights in the ($\sigma_L, \sigma_R$) polarizations. In such a framework, our recognition protocol between the $\pm$ forms consists in monitoring the time-averaged intensity $S_{3}^{ \pm}=\langle |\phi_{\rm tot}^{\pm}|^2_L - |\phi_{\rm tot}^{ \pm}|^2_R\rangle$ component of the Stokes vector associated with the total field $\phi_{\rm tot}^{\pm}= \phi_{\rm in}+ \phi_{\rm sca}^{\pm}$ transmitted behind the trap. Normalized to $\langle |\phi_{\rm in}|^2 \rangle$, 
\begin{eqnarray}
S_{3}^{ +}=(\alpha-\beta)+\frac{1}{2}\left(\alpha^2-\beta^2\right)=-S_{3}^{ -}.  \label{s3Eq}
\end{eqnarray}
The first term stems from the interference between the incident field and the scattered field. As a consequence of the conservation law of optical chirality, the scattered field is enantioselectively altered, so that the interfering term is proportional to the relative difference $\pm(\alpha-\beta)$ and hence to the circular dichroism of the single $\pm$ enantiomer. The second term $\pm(\alpha^2-\beta^2)/2$ represents the chiral field directly scattered by the trapped NPy. As such, it measures the optical chirality flux, in agreement with the prediction that optical chirality flux of opposite sign is generated by chiral objects of opposite handedness \cite{poulikakos2017inprep}. The recognition efficiency of our protocol relies in the global sign inversion of $S_3$ depending on the optically trapped $\pm$ enantiomer. For our experiments performed in the visible range, the NPys, with their pockets and tips, behave as weak light scatterers. This implies that $ |\phi_{\rm in}/ \phi_{\rm sca}^{ \pm}|\gg 1$ so that the recognition essentially operates through the dominant CD contribution. 

The experiment is depicted in detail in Fig. \ref{fig.trap}. It first immobilizes with the $785$ nm laser a single NPy enantiomer in a trapping potential made quasi-harmonic by carefully adjusting the position of the end-mirror of the SWOT. Then, a second laser, linearly polarized, is inserted inside the trap volume co-linearly with the trapping beam. This laser is slightly focused behind the trap, but to avoid exerting any force on the trapped NPy, its power is kept as low as possible ($100~\mu$W) with respect to the polarization analysis (see below). To maximize the selective dissipation of optical chirality $(\alpha-\beta)$ with respect to handedness, this second laser is tuned to the CD maximum of the NPy at $639$ nm, see Fig.~\ref{fig:CDchiral}. With a dichroic end-mirror, our configuration ensures that the $785$ nm laser is reflected, creating the SWOT, while the $639$ nm laser is perfectly transmitted by the mirror. In this way, we are able to perform the $S_3$ polarization analysis behind the trap volume by collecting, through an imaging objective (NA 0.6, $40\times$), the light transmitted and scattered in the forward direction by the NPy. The interference signal is then sent to a photodetector through a polarization analysis stage made of a quarter-wave plate at $45^\circ$, followed by a half-wave plate, and a Wollaston prism. Once the polarization analysis is performed, the NPy is released from the trap by blocking the trapping laser and the trap is re-opened after ca. $1$ min in order to catch a new NPy which is, in turn, analyzed in the same way. 

\begin{figure}[h!]
  \includegraphics[width=0.4\textwidth]{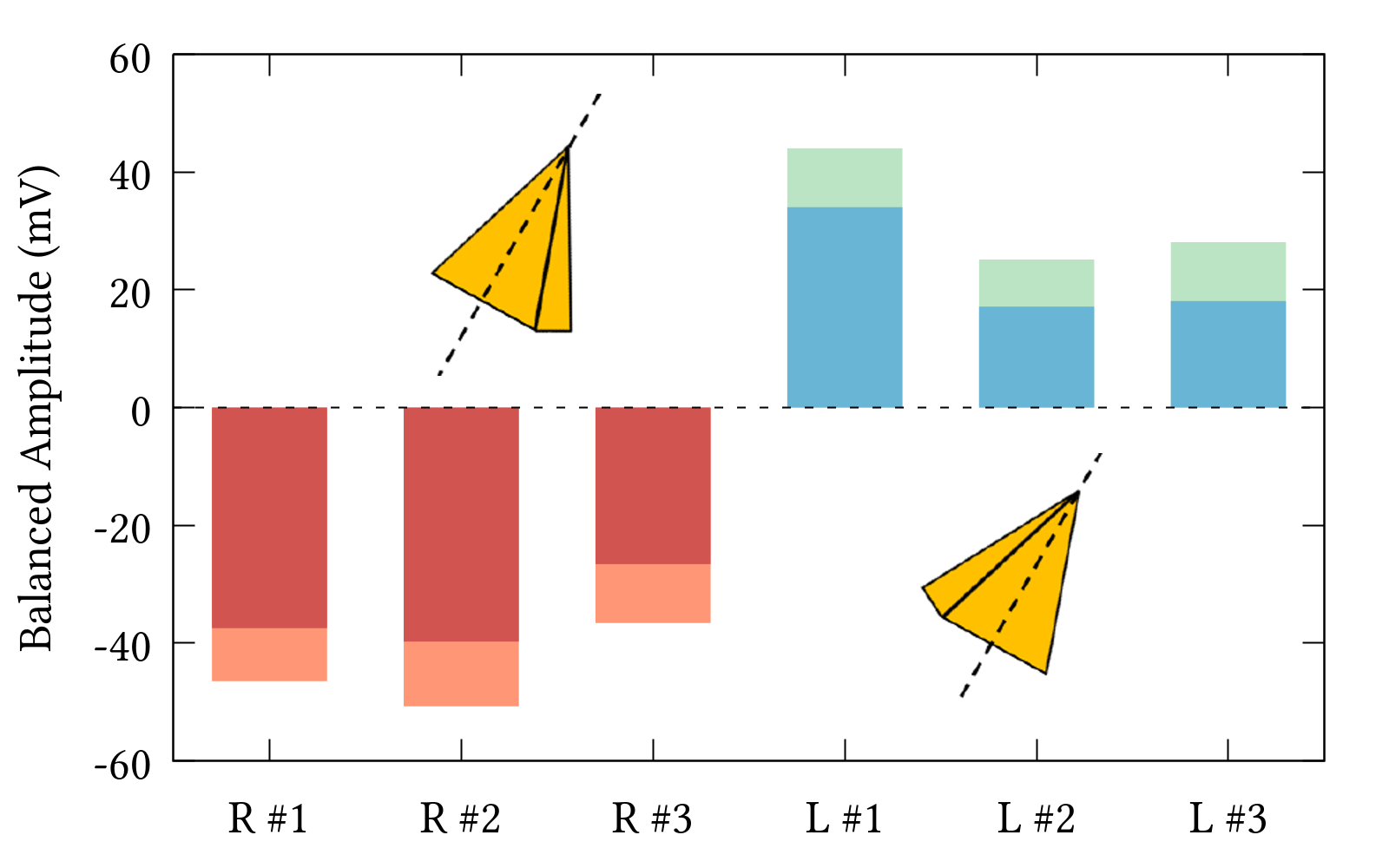}
	\caption{$S_3$ Stokes measurements for two dispersions of chiral NPys of opposite handedness. The red bars correspond to different chiral right  $+$ NPy labeled from 1 to 3 while the blue bars are 3 different left $-$ enantiomers. Errors, given by the lighter top of each bar, represent the standard deviation in measuring the $S_3$ parameter of each trapped NPy. The signal clearly exhibits non-overlapping intensity differences between the $\pm$ enantiomers. We use a fast oscilloscope to measure all $S_3$ values, averaging over an acquisition time of $\delta t=50~\mu$s. Each measurement sequence for a given dispersion is performed in less than 15 minutes and the entire comparative study was shorter than 30 minutes. These requirements are important in order to avoid fluidic drifts and NPy aggregation to affect the stability of the setup. }
	\label{fig.chirdet.trace}
\end{figure}

This procedure is repeated on two different dispersions of opposite enantiomers prepared in identical fluidic cells (identical dichroic mirrors and cover glasses) in the same way (stabilization and concentration). The two samples are analyzed in a sequential manner, following the same polarization preparation and analysis. One advantage of our experimental protocol using a Wollaston prism is that the optical settings (and in particular polarization optics) are left untouched when interchanging the fluidic cells. The measurements performed for each cell are repeated three times for validity for each $+$ and $-$ enantiomers. The single NPy trapping condition is carefully verified each time with the ISM method, and only the scattering intensities and imaging signatures corresponding to the smallest, thus single, objects are measured. The results are gathered in Fig. \ref{fig.chirdet.trace}. 

The averaged values ($\overline{S_{3}^{+}}=-39 \pm 4\ \mathrm{mV}$ and $\overline{S_{3}^{-}}=28 \pm 6\ \mathrm{mV}$) clearly show that the $+$ and $-$ enantiomeric signals can be distinguished through the polarization analysis. The reproducibility of the $S_3$ measurements for different NPys trapped from one given dispersion and within the same optical landscape suggests a constant equilibrium position of the NPys inside the optical trap. Despite this, the recorded values do not display the exact sign inversion in the $S_3$ component between the two enantiomers expected from Eq. (\ref{s3Eq}). As discussed in Supplemental Material (Sec. B and C), we explain this from $(i)$ the fact that the NPys adopt a preferred orientation inside the optical trap, and $(ii)$ from residual alignment errors in the polarization preparation and analysis stages. We show that these effects only offset the $S_3^\pm$ values by the same constant quantity, independently from the enantiomeric form. The central quantity therefore to be monitored is the difference $\Delta_{S_3} =  \overline{S_{3}^{+}}-\overline{S_{3}^{-}}$ for which the deviation from zero directly measures the NPy's preferential dissipation of incident left- or right-handed circularly polarized light, i.e. the NPy circular dichroism $\propto (\alpha-\beta)$. 

Two additional sources of variations can also be accounted for. First, small structural changes between NPys successively trapped induce distributions in the values for $\alpha,\beta$. Then, NPy thermal diffusion inside the optical trap leads to intensity variations (\textit{via} the Gaussian distribution of the trapping beam intensity) and depolarization in the forward scattering associated with an error of $\sim 5$ mV in the balanced detection for every trapped enantiomer. Such differences eventually limit the discrimination sensitivity of the experiment but despite these, our setup allows us to unambiguously measure for single NPy enantiomers $\Delta_{S_3} =-67 \pm 10$ mV, well above all data deviations. We stress once again that this result is acquired on single chiral nano-objects, optically probed while diffusing within the optical trap.

Considering the few remarkable experiments that have been performed at the micrometer scale \cite{cipparrone2011chiral,tkachenko2014helicity,dionne2017nano} or with two-dimensional objects \cite{Markovich2017preprint,Lisa2017}, our demonstration of stable optical trapping of single chiral nano-objects in three-dimensions is an important step in the development of new experimental methods for controlling and manipulating chiral nano-objects \cite{zhao2016enantioselective}. The concomitant capacity of our optical tweezer for \textit{in situ} chiral recognition gives the possibility to perform chiroptical studies on single artificial chiral objects at the nanometer scales with an unprecedented level of control. The possibility to selectively manipulate chiral matter via new modes of actuations is key for pushing the applicative potential of all-optical strategies in the vast and cross-disciplinary realm of chirality.

\paragraph{Acknowledgments} This work was supported by Agence Nationale de la Recherche (ANR), France, ANR Equipex Union (ANR-10-EQPX-52-01), the Labex NIE projects (ANR-11-LABX-0058-NIE), and the Swiss National Science Foundation under Award No. 20021-146747. Y. R.-C. is a member of the International Doctoral Program of the Initiative d'Excellence (PDI-IDEX) of the University of Strasbourg, which support is acknowledged.

\bibliography{mschiraltraparxiv}

\newpage

\section{Supplemental Material}

\section{A: Interferometric scattering microscopy}

The interferometric scattering microscopy (ISM) developed in particular by Sandoghdar \textit{et al.} \cite{Lindfors2004}, is a very efficient tool when dealing with small moving metallic scatterers. It gives us valuable informations on the quality of our metallic dispersions, allowing us to determine the nature of the trapped object, as well how many objects are actually trapped. As a label-free detection of nano-objects in fluids against a bright background, the method indeed gives the possibility to determine the size of the trapped object through the point spread function of the imaged spot and its contrast, and via the blinking dynamics of the image on the CCD video, which strongly depends on the size of the object. Small objects, with large average diffusion rates, will span the phase
through diffusion more rapidly than larger ones. To be confident that our experiments involve single NPys, we kept trapped objects only associated with the smallest ISM signals and with diffusive behavior similar to those recorded for single $150$ nm Au nanospheres. 
This has allowed us to use nanopyramid (NPy) dispersions at low concentrations, which is important for reducing the probability of having aggregates in the dispersion.

\begin{figure}[ht!]
  \centering
  \includegraphics[width=0.5\textwidth]{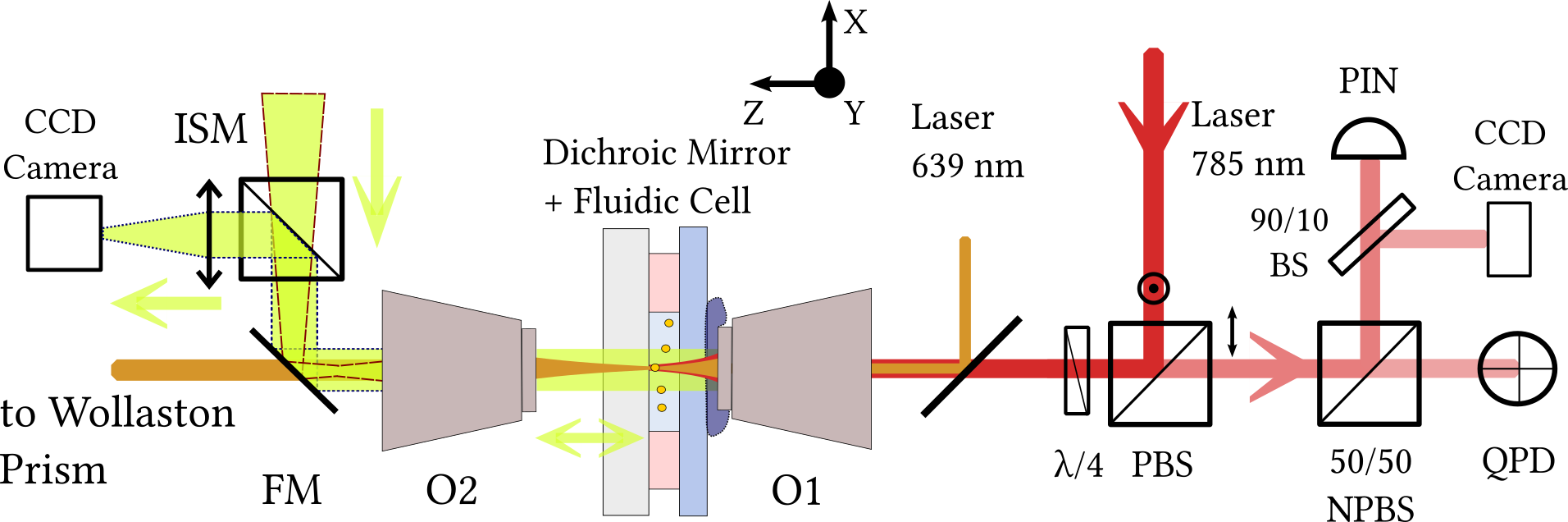}
  \caption{Schematics of the interferometric scattering spectroscopy implemented on our standing-wave optical trap (SWOT) using a flip mirror (FM). A low-power laser beam of $15~\mu$W (594 nm, 50 mW, Excelsior, Spectra Physics) is focused at the back focal plane of the objective O2 (NA 0.6, $40\times$) behind the SWOT dichroic end-mirror, transparent at 594 nm. This allows to have the laser beam almost like a plane wave between the two objectives. A fraction of this beam is scattered back at the mirror/water interface and another fraction from the NPy trapped in the fluid. The interference between these scattered beams is imaged back on a CCD camera by a tube lens. }
  \label{fig.ISM}
\end{figure}

Our ISM setup is described in Fig. \ref{fig.ISM}. We send inside the trap volume an additional low-power collimated laser and monitor on a CCD camera the interference formed between $(i)$  the partial reflection of a transmitted plane wave at the dichroic mirror / water interface and $(ii)$ the field back-scattered by the trapped NPy inside the fluidic cell.

The trapped NPy is illuminated by the additional laser beam that we assume to be close to a plane wave $E_i$. A fraction of this light is reflected at the glass/water interface forming the backward propagating reference field $E_r$. Some of the transmitted light is also back-scattered by the illuminated NPy as $\phi_s$. The two backward propagating fields $E_r$ and $\phi_s$ interfere, with a constant and dominant contribution $|E_r|^2 = |r|^2|E_i|^2$ coming from the reflection of the incident beam on the mirror's interface, characterized by a reflection amplitude $r$. A second term arises from the field $\phi_s=s\cdot E_i$ scattered by the trapped object -$s$ being the scattering amplitude. Because $|s|\ll |r|$, this term is considered as a negligible second-order term in $|s|/|r|$. The cross-term however, stemming from the interference between $E_r$ and $\phi_s$, is a first-order term that gives the relevant ISM signal. The intensity measured on the camera is then:
\begin{eqnarray}
  I_{\rm ISM} \propto |E_i|^2|r|^2(1-2\frac{|s|}{|r|}\sin\varphi),
\end{eqnarray}
with $\varphi$ the phase between the two interfering fields. The ISM approach offers a label-free detection of our NPy in the fluid against a bright background (contribution from $|r|^2|E_i|^2$). ISM provides an extremely useful technique that helps determine the size of any kind of trapped object dispersed in the fluidic cell through $(i)$ the size of the imaged spot and its contrast, and $(ii)$ through the blinking dynamics of the image in the CCD video, which strongly depends on the size of the object. Small objects, with large average diffusion rates, will span the phase $\varphi$ -a function of the distance between the object and the mirror surface- through diffusion more rapidly than larger ones. The ISM technique thus offers a straightforward way to perform our experiments. Note, however, that this simple approach only works for experiments that are not alignment sensitive. A shift of the sample position, changing the reference distances along the optical axis, can cause slight offsets between the different beams that can become detrimental. We therefore do not move any optical elements once aligned throughout the entire experiment.

\section{B: Jones matrices for chiral NPys with a fixed orientation inside the optical trap}

\begin{figure}[h!]
  \includegraphics[width=0.45\textwidth]{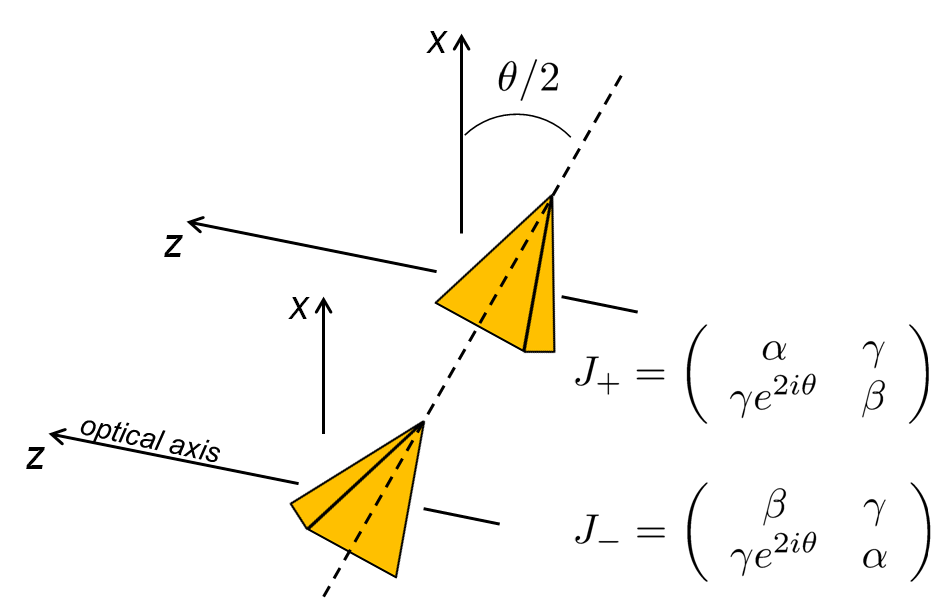}
	\caption{Schematics explaining the structure of the Jones matrices that can be associated with the $\pm$ NPy enantiomers. Due to its slight pyramidal anisotropy of shape, the $+$ enantiomer is immobilized inside the optical trap in a preferred direction that makes an angle $\theta/2$ with the $x$ axis of the optical frame -$z$ being the optical axis. The Jones matrix associated with the $-$ enantiomer is deduced by mirror symmetry.}
	\label{fig.sym}
\end{figure}

Describing the scattering on each enantiomeric form of the NPys with the simple circular dichroism (CD) Jones matrices given in Eqs. (1,2) of the main text, actually amounts to assuming some kind of rotational invariance of the chiral optical responses. This condition leads to the usual optical activity transmission matrices found for instance for isotropic chiral media such as molecular solutions. But at the single chiral object level, one has to take into account in the description the fact that the NPy takes a preferred orientation inside the optical trap due to its pyramidal shape. We observe indeed experimentally on the ISM images that each NPy adopts a stable averaged position inside the trap which corresponds, as schematized in Fig. \ref{fig.sym}, to a given orientation $\theta / 2$ taken by the base-to-tip axis of the NPy with respect to the $x$ axis inside the trap. In such conditions, one loses rotational invariance, and the Jones matrices are no longer diagonal. As discussed in detail in \cite{drezet2014reciprocity}, the Jones matrix associated with the $+$ enantiomer, for instance, must be written as
\begin{eqnarray}
J_+ = \left(\begin{array}{cc} \alpha & \gamma  \\
\gamma e^{2i\theta} & \beta  
\end{array}\right).
\end{eqnarray}
 Of course, rotational invariance imposed on such a matrix leads to set $\gamma=0$, hence recovering the simple CD case. 
  
Formally, the Jones matrix $J_-$ of the opposite enantiomeric form given in Fig. \ref{fig.sym}  is simply deduced from $J_+$ by mirror symmetry along the base-to-tip axis with 
\begin{eqnarray}
J_- = \Pi_\theta \cdot J_+\cdot\Pi_\theta^{-1}
\end{eqnarray}
where   
\begin{eqnarray}
\Pi_\theta = \left(\begin{array}{cc} 0 & e^{-i\theta}  \\
e^{i\theta} & 0 
\end{array}\right) = \Pi_\theta^{-1}
\end{eqnarray}
is the mirror symmetry transformation matrix written in the circular polarization basis. 

It is interesting to note that the Jones matrices can always be decomposed as:
\begin{eqnarray}
J_+ &=& J_{\Pi} + \frac{1}{2}(\alpha-\beta)\left(\begin{array}{cc} 1 & 0  \\
0 & -1  
\end{array}\right)  \\
J_- &=& J_{\Pi} + \frac{1}{2}(\beta-\alpha)\left(\begin{array}{cc} 1 & 0  \\
0 & -1  
\end{array}\right) 
\end{eqnarray}
where 
\begin{eqnarray}
J_{\Pi} =\left(\begin{array}{cc} (\alpha+\beta)/2 & \gamma  \\
\gamma e^{2i\theta} & (\alpha+\beta)/2  
\end{array}\right)
\end{eqnarray}
is a mirror symmetric (i.e. non-chiral) matrix \cite{drezet2014reciprocity}.

With these matrices, we now re-evaluate the time-averaged intensity $S_{3}^{ \pm}=\langle |\phi_{\rm tot}^{\pm}|^2_L - |\phi_{\rm tot}^{ \pm}|^2_R\rangle$ component of the Stokes vector associated with the total field $\phi_{\rm tot}^{\pm}= \phi_{\rm in}+ J_{\pm}\phi_{\rm in}$. To first order in $(\alpha,\beta)$, we then evaluate 
\begin{eqnarray}
S_{3}^{ +}&=&(\alpha-\beta)+ \partial  \\
S_{3}^{ -}&=&(\beta-\alpha)+ \partial   \label{ssup}
\end{eqnarray}
showing an additional constant contribution $\partial =\gamma \left(1-\cos (2\theta)\right)$ with respect to the simple CD case presented in the main text -see Eq. (3). This contribution however is consistent with our data that do not show the perfect sign inversion expected from this simple CD case. As we stressed in the main text, it is rather the difference $\Delta_{S_3} =  \overline{S_{3}^{+}}-\overline{S_{3}^{-}}$ which is meaningful in the context of optical trapped chiral objects.

Such evaluations of the Stokes vectors rely on one main assumption: the enantiomers are structural mirror images from each other, with $(\alpha,\beta,\gamma)_+=(\alpha,\beta,\gamma)_-$. This assumption cannot be absolutely true but we have indications that it is reasonable. First, as far as the $\alpha,\beta$ parameters are concerned, it is reasonable from the opposite profiles of the CD spectra associated with each enantiomers -see Fig. 1 in the main text. Then, it is also reasonable from the relatively small variations in the three successive $S_3$ measurement obtained for each $\pm$ enantiomers. Considering the reliability of the fabrication process of the NPys, in particular from an enantiomorphic point of view, the small variations observed when trapping both $\pm$ NPys imply that $(\alpha,\beta,\gamma)_+\sim(\alpha,\beta,\gamma)_-$. 

Finally, the circularly polarized trapping beam could induce a chiral optical force, as an additional radiation pressure term (see \cite{canaguier2013mechanical}). This chiral force could lead to stable positions inside the trap different for each enantiomer. Such an effect was not measured, but it would further enhance the enantiomeric separation capacity of our recognition strategy. For that reason, it is safe to consider that the positions and orientations taken by each enantiomer inside the trap are close to each other with $\theta_+\sim\theta_-$.

\section{C: Potential impact of polarization errors in preparation and analysis sequences}

\begin{figure}[h!]
  \includegraphics[width=0.48\textwidth]{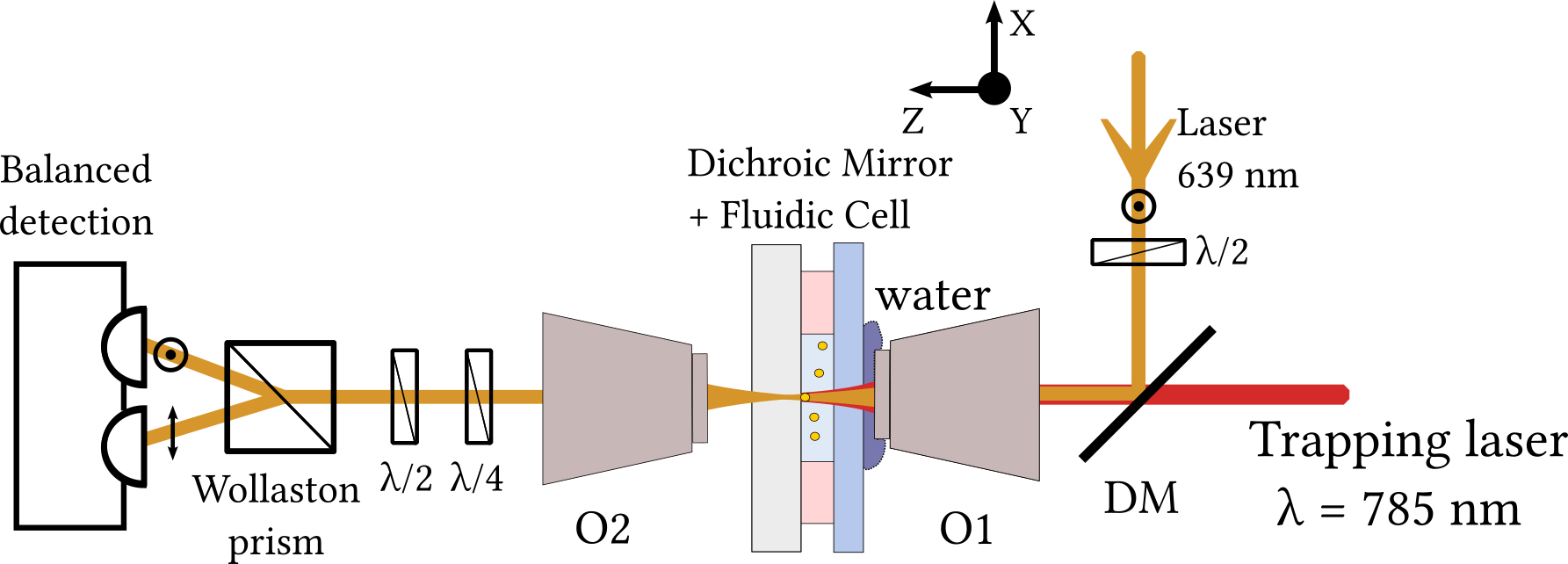}
	\caption{Polarization preparation and analysis sequences involved in the enantiomeric recognition protocol. The $45^\circ$ dichroic mirror allows injecting inside the trap the $639$ nm laser. This laser is scattered at the waist of the trapping beam by the NPy enantiomer and hence serves as the polarization probe. The preparation sequence is insured by the first half-wave plate ($\lambda/2$) and we treat the $45^\circ$ dichroic mirror - $\lambda/2$ wave plate as a phase retarding wave plate. The analysis involves a second half-wave plate ($\lambda/2$ behind the collection objective O2) and a quarter-wave plate ($\lambda/4$) which fast axes are respectively set to $\pi/2+\theta_{\lambda/2}$ and $\pi/4+\theta_{\lambda/4}$. }
	\label{fig.zoom}
\end{figure}

Our polarimetry takes the advantage of using a Wollaston prism that analyzes left vs. right handed circular polarized $\sigma_+,\sigma_-$ states through a mere intensity balanced detection. It hence avoids having to manipulate and rotate any wave plate during the analysis. For an empty trap, the half-wave plate ($\lambda/2$) placed before the prism (but after the collection objective, as shown in Fig. \ref{fig.zoom}) is adjusted to precisely compensate slight misalignments between the prism and the balanced detector. When set, the $\lambda/2$ wave plate yields zero in the balanced detection for an linearly polarized input state and absolute maximum (minimum) for $\sigma_+$ ($\sigma_-$). In the analysis sequence therefore, the main source of errors, however very small, will come from the alignment of the fast axis of the quarter-wave plate ($\lambda/4$).

In the preparation sequence in contrast, the presence of a $45^\circ$ dichroic mirror (DM in Fig. \ref{fig.zoom}) induces polarization errors that have to be discussed carefully. Because the reflection amplitudes for $s$ and $p$ polarizations are different on DM, we use a motorized half-wave plate to set the input linear polarization of the $639$ nm laser normal to the plane of incidence of the beam on the $45^\circ$ dichroic mirror, i.e. as close to the vertical $y$ axis as possible (see frame on Fig. \ref{fig.zoom}). This corresponds to an orientation of the field $\vartheta = \pi/2 + \delta$ of the half-wave plate fast axis, but with an unavoidable small offset $\delta$. Then, in order to keep the discussion of polarization errors general, we simply model the $\lambda/2$-DM system as a general phase retarding wave plate with $\tilde{J}(\pi/2+\delta, \eta, \varphi)$, where $\eta$ is the relative phase retardation between the fast and slow axes, and $\varphi$ the circular retardance (following the conventions of \cite{goldstein2017polarized}). 

The polarization analysis sequence can then be written in a straightforward way. We start with the incident field $\phi_{\rm in}$ linearly polarized, first sent through the half-wave plate-DM system as $\tilde{\phi}_{\rm in}=\tilde{J}(\pi/2+\delta, \eta, \varphi)\phi_{\rm in}$. This field then illuminates the chiral sample inside the trap and is transmitted as $\tilde{\phi}_{\rm tot}^{\pm}=\tilde{\phi}_{\rm in}+J^{\pm}\tilde{\phi}_{\rm in}$. As explained in the main text, we measure the $S_{3}^{\pm}$ parameters from a balanced detection of intensities $\Delta I$ in the horizontal $|H\rangle$ and vertical $|V\rangle$ states of polarization after the Wollaston prism.  This balanced detection yields:
\begin{eqnarray}
\Delta I&=&|\langle H | J_{\frac{\lambda}{4}}(\pi/4+\theta_{\lambda/4}) \cdot J_{\frac{\lambda}{2}}(\pi/2+\theta_{\lambda/2}) | \tilde{\phi}_{\rm tot}^{\pm}\rangle |^2  \\ \nonumber 
&&-|\langle V | J_{\frac{\lambda}{4}}(\pi/4 +\theta_{\lambda/4})\cdot J_{\frac{\lambda}{2}}(\pi/2+\theta_{\lambda/2}) | \tilde{\phi}_{\rm tot}^{\pm}\rangle |^2, \nonumber 
\end{eqnarray}
where $J_{\frac{\lambda}{4}}(\pi/4+\theta_{\lambda/4})$ and $J_{\frac{\lambda}{2}}(\pi/2+\theta_{\lambda/2})$ are the Jones matrices associated with the analysis quarter- and half-wave plates. They include in $\theta_{\lambda/4}$ and $\theta_{\lambda/2}$ small deviations from the perfect $\pi/4$ and $\pi/2$ orientations of the fast axes. 

Due to these deviations and the $\lambda/2$-DM system, we are confronted to a residual contribution $\delta I$ from the direct transmission that is not perfectly canceled in the balanced detection. This residual contribution can be expanded to second order in the errors as
\begin{eqnarray}
\delta I \sim - 2 (8\delta  \eta \theta_{\lambda/4}\theta_{\lambda/2}-4\theta_{\lambda/4}\theta_{\lambda/2} +\delta\eta)
\end{eqnarray}

This additive term, which does not depend on the enantiomeric form of the NPy, therefore acts exactly on the same level as the orientational effect discussed above: it forbids to measure the expected exact sign inversion between the $S_{3}^{+}$ and $S_{3}^{-}$ parameters. But as discussed above, it can be eliminated by measuring the difference in the Stokes parameters for the two enantiomers. From the polarization analysis point of view, this difference then only depends on the relative orientation of the wave plates. The difference can be derived at the second-order in potential polarization misalignment errors (and at the first order in the $(\alpha,\beta)$ chiral response) as:
\begin{eqnarray}
S_{3}^{+}-S_{3}^{-} \sim (\alpha -\beta) \cdot \left(1+8 \theta_{\lambda/4} \theta_{\lambda/2}\right).
\end{eqnarray}

We emphasize that the $(\pi/4, \pi/2)$ orientations of the fast axes of the analysis quarter- and half-wave plates are actually the best controlled parameters of the entire polarimetric protocol. This implies that the angular deviations $(\theta_{\lambda/4} ,\theta_{\lambda/2})$ are much too small to change the overall sign of the $S_{3}^{+}-S_{3}^{-}$ difference. We can hence safely conclude that this difference is robust to polarization errors both in the preparation and in the analysis sequences, providing for that reason a reliable observable for recognizing the two different NPy enantiomers.

\end{document}